\documentclass[pdflatex]{sn-jnl}
\jyear{2022}%
\theoremstyle{thmstyleone}%
\theoremstyle{thmstyletwo}%
\theoremstyle{thmstylethree}%
\usepackage{tabularx}
\usepackage{hyperref}
\usepackage{diagbox}
\usepackage{threeparttable}
\usepackage{setspace}
\usepackage{indentfirst}
\usepackage[superscript]{cite}

\usepackage{lineno}

\newcounter{firstbib}
\begin{document}
\title{Vela pulsar wind nebula x-rays are polarized to near the synchrotron limit}
\author{
Fei Xie$^{1,2}$\thanks{Corresponding author: xief@gxu.edu.cn},
Alessandro {Di Marco}$^{2}$,
Fabio {La Monaca}$^{2}$,
Kuan Liu$^{1}$,
Fabio Muleri$^{2}$,
Niccol\`{o} Bucciantini$^{3,4,5}$,
Roger W. Romani$^{6}$,
Enrico Costa$^{2}$,
John Rankin$^{2}$,
Paolo Soffitta$^{2}$,
Matteo Bachetti$^{7}$,
Niccol\`{o} Di Lalla$^{6}$,
Sergio Fabiani$^{2}$,
Riccardo Ferrazzoli$^{2}$,
Shuichi Gunji$^{8}$,
Luca Latronico$^{9}$,
Michela Negro$^{10,11,12}$,
Nicola Omodei$^{6}$,
Maura Pilia$^{7}$,
Alessio Trois$^{7}$,
Eri Watanabe$^{8}$,
{Iv\'an} Agudo$^{13}$,
Lucio A. Antonelli$^{14,15}$,
Luca Baldini$^{16,17}$,
Wayne H. Baumgartner$^{18}$,
Ronaldo Bellazzini$^{16}$,
Stefano Bianchi$^{19}$,
Stephen D. Bongiorno$^{18}$,
Raffaella Bonino$^{9,20}$,
Alessandro Brez$^{16}$,
Fiamma Capitanio$^{2}$,
Simone Castellano$^{16}$,
Elisabetta Cavazzuti$^{21}$,
Stefano Ciprini$^{15,22}$,
Alessandra De Rosa$^{2}$,
Ettore Del Monte$^{2}$,
Laura Di Gesu$^{21}$,
Immacolata Donnarumma$^{21}$,
Victor Doroshenko$^{23}$,
Michal Dovčiak$^{24}$,
Steven R. Ehlert$^{18}$,
Teruaki Enoto$^{25}$,
Yuri Evangelista$^{2}$,
Javier A. Garcia$^{26}$,
Kiyoshi Hayashida$^{27,\dagger}$,
Jeremy Heyl$^{28}$,
Wataru Iwakiri$^{29}$,
Svetlana G. Jorstad$^{30,31}$,
Vladimir Karas$^{24}$,
Takao Kitaguchi$^{25}$,
Jeffery J. Kolodziejczak$^{18}$,
Henric Krawczynski$^{32}$,
Ioannis Liodakis$^{33}$,
Simone Maldera$^{9}$,
Alberto Manfreda$^{16}$,
Fr\'{e}d\'{e}ric Marin$^{34}$,
Andrea Marinucci$^{21}$,
Alan P. Marscher$^{30}$,
Herman L. Marshall$^{35}$,
Francesco Massaro$^{9,20}$,
Giorgio Matt$^{19}$,
Ikuyuki Mitsuishi$^{36}$,
Tsunefumi Mizuno$^{37}$,
C.-Y. Ng$^{38}$,
Stephen L. O'Dell$^{18}$,
Chiara Oppedisano$^{9}$,
Alessandro Papitto$^{14}$,
George G. Pavlov$^{39}$,
Abel L. Peirson$^{6}$,
Matteo Perri$^{14,15}$,
Melissa Pesce-Rollins$^{16}$,
Pierre-Olivier Petrucci$^{40}$,
Andrea Possenti$^{7}$,
Juri Poutanen$^{41,42}$,
Simonetta Puccetti$^{21}$,
Brian D. Ramsey$^{18}$,
Ajay Ratheesh$^{2}$,
Carmelo Sgr\'{o}$^{16}$,
Patrick Slane$^{43}$,
Gloria Spandre$^{16}$,
Toru Tamagawa$^{25}$,
Fabrizio Tavecchio$^{44}$,
Roberto Taverna$^{45}$,
Yuzuru Tawara$^{36}$,
Allyn F. Tennant$^{18}$,
Nicolas E. Thomas$^{18}$,
Francesco Tombesi$^{46}$,
Sergey S. Tsygankov$^{41,42}$,
Roberto Turolla$^{45,47}$,
Jacco Vink$^{48}$,
Martin C. Weisskopf$^{18}$,
Kinwah Wu$^{47}$,
Silvia Zane$^{47}$
\begin{spacing}{0.7}
*Corresponding author: xief@gxu.edu.cn
\end{spacing}
}

\affil[1]{Guangxi Key Laboratory for Relativistic Astrophysics, School of Physical Science and Technology, Guangxi University, Nanning 530004, China}
\affil[2]{INAF Istituto di Astrofisica e Planetologia Spaziali, Via del Fosso del Cavaliere 100, 00133 Roma, Italy}
\affil[3]{INAF Osservatorio Astrofisico di Arcetri, Largo Enrico Fermi 5, 50125 Firenze, Italy}
\affil[4]{Dipartimento di Fisica e Astronomia, Universit\`{a} degli Studi di Firenze, Via Sansone 1, 50019 Sesto Fiorentino (FI), Italy} 
\affil[5]{Istituto Nazionale di Fisica Nucleare, Sezione di Firenze, Via Sansone 1, 50019 Sesto Fiorentino (FI), Italy} 
\affil[6]{Department of Physics and Kavli Institute for Particle Astrophysics and Cosmology, Stanford University, Stanford, California 94305, USA}
\affil[7]{INAF Osservatorio Astronomico di Cagliari, Via della Scienza 5, 09047 Selargius (CA), Italy}
\affil[8]{Yamagata University,1-4-12 Kojirakawa-machi, Yamagata-shi 990-8560, Japan}
\affil[9]{Istituto Nazionale di Fisica Nucleare, Sezione di Torino, Via Pietro Giuria 1, 10125 Torino, Italy} 
\affil[10]{NASA Goddard Space Flight Center; Greenbelt, MD 20771, USA}
\affil[11]{University of Maryland, Baltimore County, Baltimore, MD 21250, USA}
\affil[12]{Center for Research and Exploration in Space Science and Technology, NASA/GSFC, Greenbelt, MD 20771, USA}
\affil[13]{Instituto de Astrof\'isica de Andaluc\'ia—CSIC, Glorieta de la Astronom\'ia s/n, 18008, Granada, Spain}
\affil[14]{INAF Osservatorio Astronomico di Roma, Via Frascati 33, 00040 Monte Porzio Catone (RM), Italy} 
\affil[15]{Space Science Data Center, Agenzia Spaziale Italiana, Via del Politecnico snc, 00133 Roma, Italy} 
\affil[16]{Istituto Nazionale di Fisica Nucleare, Sezione di Pisa, Largo B. Pontecorvo 3, 56127 Pisa, Italy}  
\affil[17]{Dipartimento di Fisica, Universit\`{a} di Pisa, Largo B. Pontecorvo 3, 56127 Pisa, Italy} 
\affil[18]{NASA Marshall Space Flight Center, Huntsville, AL 35812, USA} 
\affil[19]{Dipartimento di Matematica e Fisica, Universit\`{a} degli Studi Roma Tre, Via della Vasca Navale 84, 00146 Roma, Italy} 
\affil[20]{Dipartimento di Fisica, Universit\`{a} degli Studi di Torino, Via Pietro Giuria 1, 10125 Torino, Italy}
\affil[21]{Agenzia Spaziale Italiana, Via del Politecnico snc, 00133 Roma, Italy}
\affil[22]{Istituto Nazionale di Fisica Nucleare, Sezione di Roma Tor Vergata, Via della Ricerca Scientifica 1, 00133 Roma, Italy}
\affil[23]{Institut f\"{u}r Astronomie und Astrophysik, Sand 1, 72076 T\"{u}bingen, Germany}
\affil[24]{Astronomical Institute of the Czech Academy of Sciences, Bočn\'{i} II 1401/1, 14100 Praha 4, Czech Republic} 
\affil[25]{RIKEN Cluster for Pioneering Research, 2-1 Hirosawa, Wako, Saitama 351-0198, Japan} 
\affil[26]{California Institute of Technology, Pasadena, CA 91125, USA} 
\affil[27]{Osaka University, 1-1 Yamadaoka, Suita, Osaka 565-0871, Japan}
\affil[28]{University of British Columbia, Vancouver, BC V6T 1Z4, Canada}
\affil[29]{Department of Physics, Faculty of Science and Engineering, Chuo University, 1-13-27 Kasuga, Bunkyo-ku, Tokyo 112-8551, Japan}
\affil[30]{Institute for Astrophysical Research, Boston University, 725 Commonwealth Avenue, Boston, MA 02215, USA}
\affil[31]{Laboratory of Observational Astrophysics, St. Petersburg University, University Embankment 7/9, St. Petersburg 199034, Russia}
\affil[32]{Physics Department and McDonnell Center for the Space Sciences, Washington University in St. Louis, St. Louis, MO 63130, USA}
\affil[33]{Finnish Centre for Astronomy with ESO,  Vesilinnantie 5, 20014 University of Turku, Finland}
\affil[34]{Universit\'{e} de Strasbourg, CNRS, Observatoire Astronomique de Strasbourg, UMR 7550, 67000 Strasbourg, France}
\affil[35]{MIT Kavli Institute for Astrophysics and Space Research, Massachusetts Institute of Technology, 77 Massachusetts Avenue, Cambridge, MA 02139, USA}
\affil[36]{Graduate School of Science, Division of Particle and Astrophysical Science, Nagoya University, Furo-cho, Chikusa-ku, Nagoya, Aichi 464-8602, Japan}
\affil[37]{Hiroshima Astrophysical Science Center, Hiroshima University, 1-3-1 Kagamiyama, Higashi-Hiroshima, Hiroshima 739-8526, Japan}
\affil[38]{Department of Physics, University of Hong Kong, Pokfulam, Hong Kong}
\affil[39]{Department of Astronomy and Astrophysics, Pennsylvania State University, University Park, PA 16801, USA}
\affil[40]{Universit\'{e} Grenoble Alpes, CNRS, IPAG, 38000 Grenoble, France}
\affil[41]{Department of Physics and Astronomy, University of Turku, 20014 Turku, Finland}
\affil[42]{Space Research Institute of the Russian Academy of Sciences, Profsoyuznaya Str. 84/32, Moscow 117997, Russia}
\affil[43]{Center for Astrophysics, Harvard \& Smithsonian, 60 Garden St, Cambridge, MA 02138, USA}
\affil[44]{INAF Osservatorio Astronomico di Brera, via E. Bianchi 46, 23807 Merate (LC), Italy}
\affil[45]{Dipartimento di Fisica e Astronomia, Universit\`{a} degli Studi di Padova, Via Marzolo 8, 35131 Padova, Italy}
\affil[46]{Dipartimento di Fisica, Universit\`{a} degli Studi di Roma Tor Vergata, Via della Ricerca Scientifica, 00133 Roma, Italy}
\affil[47]{Mullard Space Science Laboratory, University College London, Holmbury St Mary, Dorking, Surrey RH5 6NT, UK}
\affil[48]{Anton Pannekoek Institute for Astronomy \& GRAPPA, University of Amsterdam, Science Park 904, 1098 XH Amsterdam, The Netherlands}
\maketitle
\pagestyle{plain}

 \textbf{Pulsar wind nebulae are formed when outflows of relativistic electrons and positrons hit the surrounding supernova remnant or interstellar medium at a shock front. The Vela pulsar wind nebula is powered by a young pulsar (B0833-45, age 11 kyr) \cite{1968Natur.220..340L} and located inside an extended structure called Vela X, itself inside the supernova remnant \cite{1985ApJ...299..828H}\!. Previous X-ray observations revealed two prominent arcs, bisected by a jet and counter jet \cite{2001ApJ...554L.189P,2001ApJ...556..380H}\!. Radio maps have shown high linear polarization of 60 per cent in the outer regions of the nebula \cite{2003MNRAS.343..116D}\!. Here we report X-ray observation of the inner part of the nebula, where polarization can exceed 60 per cent at the leading edge, which approaches the theoretical limit of what can be produced by synchrotron emission. We infer that, in contrast with the case of the supernova remnant, the electrons in the pulsar wind nebula are accelerated with little or no turbulence in a highly uniform magnetic field.}
\\

\,
Imaging X-ray Polarimetry Explorer (IXPE) \cite{2021AJ....162..208S,2022JATIS...8b6002W}\! provides a new X-ray view of the highest energy electrons near their acceleration sites. IXPE is a NASA/ASI explorer featuring three co-aligned X-ray telescopes each with an imaging photoelectric polarimeter detector unit based on a gas pixel detector \cite{Costa2001,2021APh...13302628B}\!. IXPE observed the Vela pulsar wind nebula (PWN) in two periods: (i) 2022-04-05 to 2022-04-15 and (ii) 2022-04-21 to 2022-04-30, with a total exposure of 860\,ks. The data were extracted from the publicly available files processed by the IXPE Science Operations Center and analysed with standard tools as described in the supplementary Methods section.

Linear polarization is detected at high significance ($\sim$\,31$\sigma$) for the spatial- and energy-integrated Vela PWN in the 2--8\,keV band. 
The linear polarization degree (PD) is (44.6$\pm$1.4)\% with a polarization angle (PA, also known as electric vector position angle EVPA, defined from North through East) of $-(50.0\pm$0.9)$^\circ$, with 68.3\% confidence level uncertainties. Previously, only the Crab PWN has been studied via X-ray polarization; the classic 1976, 1978 OSO-8 results \cite{1976ApJ...208L.125W,1978ApJ...220L.117W} have recently been confirmed by the Polarlight cube-sat \cite{2020NatAs...4..511F}\!. However, these observations provided only an integrated polarization measurement and the image-averaged polarization degree was less than half of that found here for Vela.

\begin{figure}
\centering
\includegraphics[width=1.\textwidth]{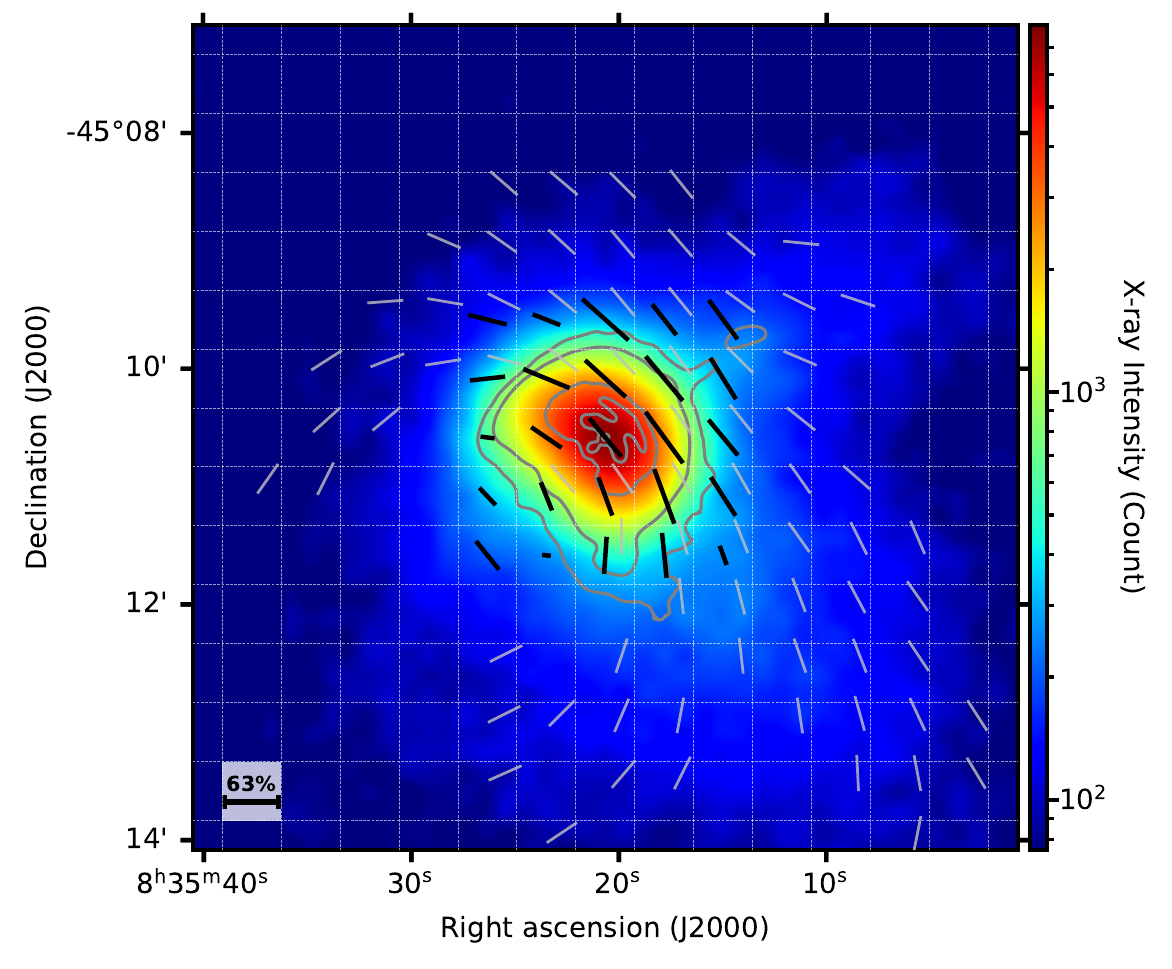}
\caption{
The IXPE intensity map of the Vela PWN in the 2--8\,keV range with the measured X-ray polarization and radio polarization vectors overlaid. Intensity is the Gaussian-smoothed sum of the three detector units. The black lines present the X-ray polarization in the corresponding 30$^{\prime\prime}\times$30$^{\prime\prime}$ image elements (white grids), their lengths indicate the polarization degree and the orientation indicates the projected magnetic field ($90^\circ$ from the EVPA). 
The thinner silver lines, drawn with the same lengths, show the 5 GHz polarization directions derived from ref. \cite{2003MNRAS.343..116D}\!. The gray contours, obtained from Chandra observations in the same 2--8\,keV range, give a better illustration of the pulsar, compact arc-jet structure, diffuse emission shell, and relatively weak outer jet (only weakly detected at IXPE resolution). The bar at the bottom left presents the maximum measured X-ray polarization degree of 63\%, and the PD and PA in each of 25 grids are tabulated in Extended Data Table \ref{tab:pol_map_3dus}.}
\label{fig:pol_map}
\end{figure}

IXPE's imaging capabilities, with a $\le 30^{\prime\prime}$ half power diameter \cite{2022JATIS...8b6002W}\!, allow a spatially resolved polarimetric measurement of the PWN (Fig.\,\ref{fig:pol_map}). Here, 25 independent 30$^{\prime\prime}\times$30$^{\prime\prime}$ square regions have been analyzed, combining data from the 3 detector units. The pulsar and compact arc-jet structure lie largely within the central region. The pulsar, dominated by a $kT \sim 0.13$\,keV thermal component \cite{2001ApJ...552L.129P}\!, contributes less than 10\% of the IXPE counts in this region. IXPE has not measured the polarization of the pulsed source; this may add to or subtract from the nebular polarization, but with the low flux (and low expected polarization degree \cite{Bucciantini et al}) the effect should be small. The black lines show linear polarization, with line length indicating polarization degree up to 62.8\%. These lines are at 90$^\circ$ to the EVPA, and thus show the direction of the projected magnetic field, which is highly symmetric about the pulsar jet axis.

The curved and symmetric polarization angle pattern seen in Fig.\,\ref{fig:pol_map} implies that there will be some polarization angle variation across, and decrease in the region-average polarization degree of, our 25 measurement regions. This is most obvious to the sides and rear of the PWN, where the arcs seen in the Chandra image are highly curved and thus the underlying polarization angle varies rapidly, producing substantial de-polarization.  
Since the polarization follows the Chandra X-ray morphology we expect slightly higher polarization degree and smaller errors in regions where the arcs are less curved, which allows us to average over larger areas. As an example a morphologically selected region at the front of the nebula, along the symmetry axis provides a PD=70.0$\pm$3.6\% (Methods and Extended Data Fig.\,\ref{fig:panda_region}).
As a caution, we note that for an analysis near the angular resolution limit in regions covering sharp intensity gradients, an effect due to the reconstruction of the photo-electron tracks in the gas pixel detector can artificially alter the local polarization values. Monte Carlo analysis show that this is a $\lesssim 5$\% effect, with a slight decrease in polarization degree to the North and increase to the South; our maximally polarized zones are hardly affected.
Regardless, Vela's high polarization degree and symmetric pattern imply a highly ordered magnetic field following the PWN's toroidal structure.

The spatially-averaged polarization of the Vela PWN is plotted as a function of energy in Fig.\,\ref{fig:pol_energy}. A slight increase of polarization degree with energy is seen. This might be expected, since the emission region shrinks with increasing energy, thus arising from a smaller range of magnetic field orientations. The image-averaged polarization angle does not show significant energy dependence.

It is widely accepted that PWN arc-jet structures are due to an anisotropic pulsar wind, which generates toroidal magnetic fields in the equatorial zone \cite{2004ApJ...601..479N,2002MNRAS.329L..34L,2003MNRAS.344L..93K,2004A&A...421.1063D}\!.
In this picture, the integrated polarization angle should align with the symmetry axis of the Vela PWN \cite{2004MNRAS.349..779K,2005A&A...443..519B,2006A&A...453..621D}\!, well consistent with the results reported here.
The non-thermal radio and X-ray spectra of the Vela PWN already indicated domination by synchrotron emission. The high linear X-ray polarization, measured here for the first time with IXPE, strengthens this conclusion. 

The maximum polarization degree of synchrotron radiation from a power law spectrum of electrons with index $p$ in a uniform magnetic field is $\Pi=(p+1)/(p+7/3)$ \cite{1986rpa..book.....R}\!;
the photon index $\Gamma$ of the emitted radiation is given by $p=2\Gamma-1$. For Vela, with $\Gamma = 1.3\pm 0.04$ in the compact jet-arc structure, increasing to 1.7 in the outskirts \cite{2002ASPC..271..181K}\!, the maximum polarization degree is 66--72\%. Thus the 62.8\% seen in Fig.\,\ref{fig:pol_map} and Extended Data Table\,\ref{tab:pol_map_3dus} (or the 70\% in regions selected according to the Chandra image in Extended Data Table\,\ref{tab:panda}) is quite close to the maximum permitted value. This implies a magnetic field that is highly uniform across the measured region, with little turbulence-induced fluctuations. 
Also $\Gamma=1.3$ implies $p=1.6$, appreciably flatter than the $p\approx 2.3-2.6$ of turbulent diffusive shock acceleration \cite{2015SSRv..191..519S}\!; another mechanism, such as reconnection \cite{2014ApJ...783L..21S}\!, should play a dominant role in PWN particle energization.

The high uniformity and toroidal magnetic structure that we see with IXPE evidently extends to larger radii, where the $4'$ radio lobes' linear polarization indicates a magnetic field with a similar symmetry axis \cite{2003MNRAS.343..116D} (Fig.\,\ref{fig:pol_map}). Interestingly this magnetic structure is compressed to the North (evidently by the PWN interaction with the larger supernova remnant) and has a larger radius of curvature, centered behind the pulsar. With little radio emission from the X-ray bright torus/jet zone in the center, it appears that the radio tracks an older, cooler electron population centered behind the pulsar and extending to larger radii. Still, the substantial radio polarization, reaching 60\% at both 5-GHz \cite{2003MNRAS.343..116D} and 1.4-GHz \cite{2002ASPC..271..187B}\!, indicates that the ordered and minimally turbulent fields probed by IXPE extend into this radio-emitting zone at much larger radii.

At present, we can say little about the X-ray polarization from the pulsar itself. Much deeper observations (and likely higher spatial resolution) will be required to isolate this component and compare with the optical phase average polarization of PD=8.1$\pm0.6$\% at PA=146.3$\pm$2.4$^\circ$\cite{2014MNRAS.445..835M}\!. Like the Crab pulsar, Vela's average optical polarization angle lies close to the projected torus axis. If, like the Crab pulsar, the phase-average X-ray polarization degree is well below that of the optical emission, corrections to our nebular estimates will be very small.

\bigskip
We find a remarkable high X-ray polarization in the Vela PWN, reaching an image- and energy-averaged polarization degree of $\sim$\,45\%. The non-thermal PWN spectrum is bright in the IXPE energy band; the Vela pulsar itself emits mostly soft thermal X-rays, too faint and weakly polarized for detection at present. Our IXPE image sufficiently resolves the PWN to show that the polarization structure is symmetric about the projected pulsar spin (and proper motion) axis; this symmetry extends in radio polarization studies to even larger angles. The varying polarization angle across the nebula implies that the true local polarization is even larger and indeed we find polarization degree values $\geq$\,60\% for some regions. This is close to the maximum polarization degree allowed for synchrotron emission with the observed X-ray spectrum, implying that the magnetic field is highly uniform across the emission region. 
Since electrons emitting synchrotron X-rays cool very rapidly, the IXPE X-rays are emitted close to the acceleration zone. In turn this argues against turbulence-driven diffusive shock acceleration and suggests that other processes, such as reconnection, should energize the PWN particles in the termination shock.
Further IXPE studies of Vela and other bright PWNe should connect the X-ray polarization pattern with the details of the compact structures, further probing the physics of relativistic shock acceleration.

\clearpage

\clearpage
\clearpage
\begin{figure}
\centering
\includegraphics[width=0.8\textwidth]{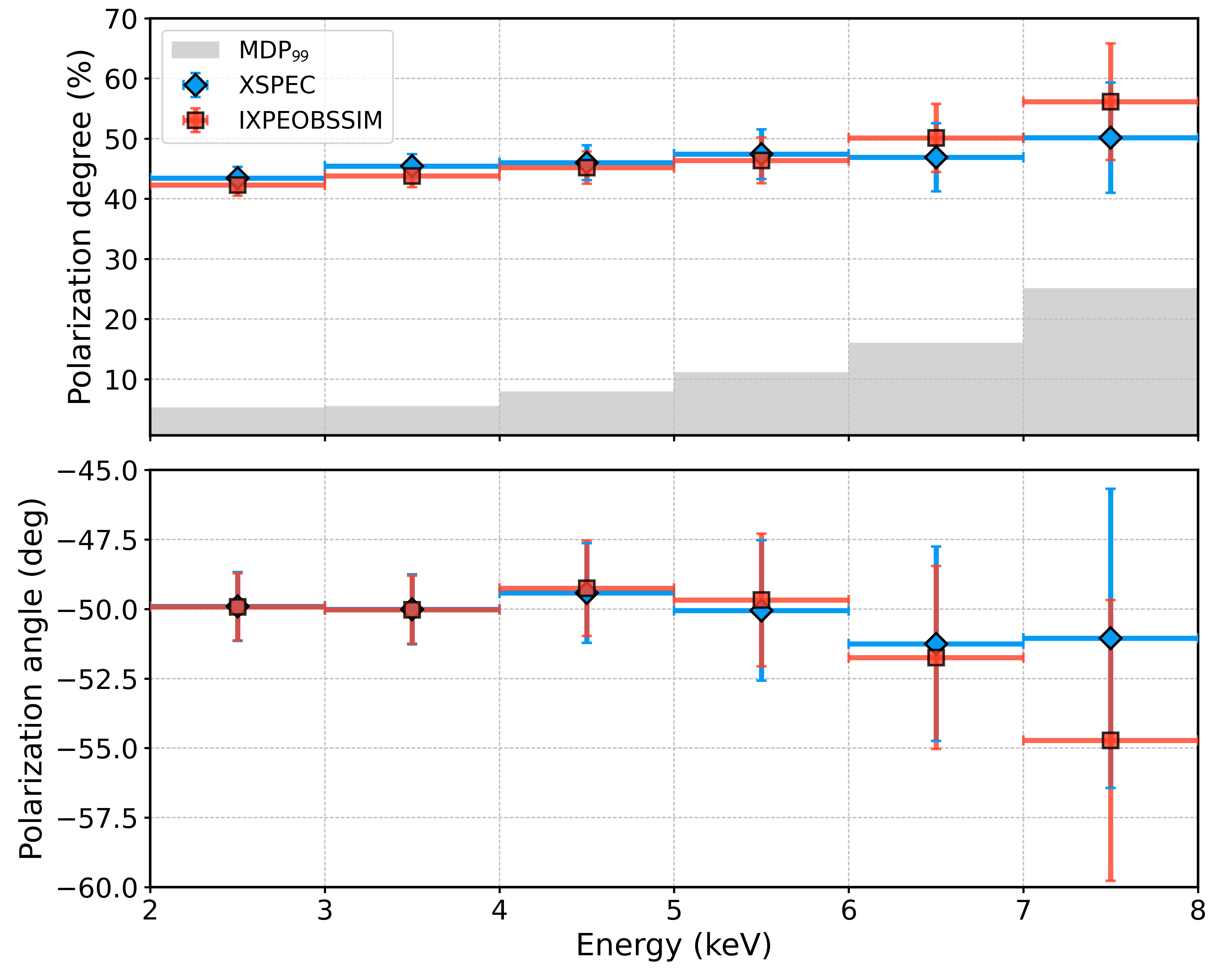}
\caption{Image-average polarization of the Vela PWN as a function of energy. Polarization degree (top) and polarization angle (bottom) are derived with \textit{ixpeobssim} (red) and \textsc{xspec} (blue) independently. The results are from the joint analysis of three detector units with uncertainties calculated for a 68.3\% confidence level.
Polarization degrees in all energy bins are significantly higher than the minimum detectable polarization at the 99\% confidence level (MDP$_{99}$) shown as a gray histogram.}
\label{fig:pol_energy}
\end{figure}

\clearpage
\section*{Methods}

\subsection*{IXPE Data Analysis}

This work is based on the polarimetric observations of the Vela pulsar and nebula obtained with the Imaging X-ray Polarimetry Explorer (IXPE), a NASA mission in partnership with the Italian space agency (ASI), as described in ref. \cite{2021AJ....162..208S,2022JATIS...8b6002W} and references therein. The IXPE Observatory includes three identical X-ray telescopes, each comprising an X-ray mirror assembly and a linear-polarization-sensitive detector, and performs imaging polarimetry over a nominal 2--8\,keV energy band. IXPE data are telemetered to ground stations in Malindi (primary) and in Singapore (secondary), and are transmitted to the NASA Marshall Space Flight Center Science Operations Center (SOC) via the Mission Operations Center (MOC), at the Laboratory for Atmospheric and Space Physics, University of Colorado. Data are processed at the SOC, where raw data and relevant engineering/ancillary data are used to produce photon event lists. For each observation, the data are archived at the High-Energy Astrophysics Science Archive Research Center (HEASARC), at the NASA Goddard Space Flight Center for use by the international astrophysics community.

IXPE data are provided in two levels/formats: event\_l1 and event\_l2. The first level consists of unfiltered events; event\_l2 files are produced from event\_l1 after further calibration/correction procedures. In particular, event\_l2 data are cleaned of in-flight-calibration data and occultation/SAA passages, then energy calibration/equalization of the three detector units (DUs) is applied and Stokes parameters are estimated following ref. \cite{2015APh....68...45K}\!. After this the spurious modulation is removed \cite{Rankin_2022} and the photons are referenced to sky coordinates, removing dithering pattern and boom motion effects.

The Vela pulsar and PWN were observed with IXPE from 2022-04-05 at 19:51:40.687 UTC to 2022-04-15 at 18:08:18.343 UTC, and again from 2022-04-21 at 12:22:11.549 UTC to 2022-04-30 at 10:34:51.991 UTC, with net exposures 860\,ks, 854\,ks, 868\,ks, respectively for the three DUs. After retrieving event\_l2 files from the HEASARC, we performed three corrections: (1) energy correction (based on the onboard calibration data); (2) bad time interval filtering (to remove the solar flare events, see Extended Data Fig.\,\ref{fig:solar_remove}); (3) barycenter correction (to convert the photon arrival time into the solar system barycenter; done with the standard HeaSoft 6.30.1 \textit{barycorr}, with the following options: refframe=ICRS, ephem=JPLEPH.421). 
We find that the IXPE Vela observations were slightly affected by an increased particle flux during some portions of the orbit, associated with high  solar activity.

Bad time intervals are identified as periods with count rates 3$\sigma$ higher than average, assuming a Gaussian distribution. Events during such bad intervals are excluded from further data analysis.

\begin{figure}
\renewcommand{\figurename}{Extended Data Fig.} 
\centering
\includegraphics[width=1\textwidth]{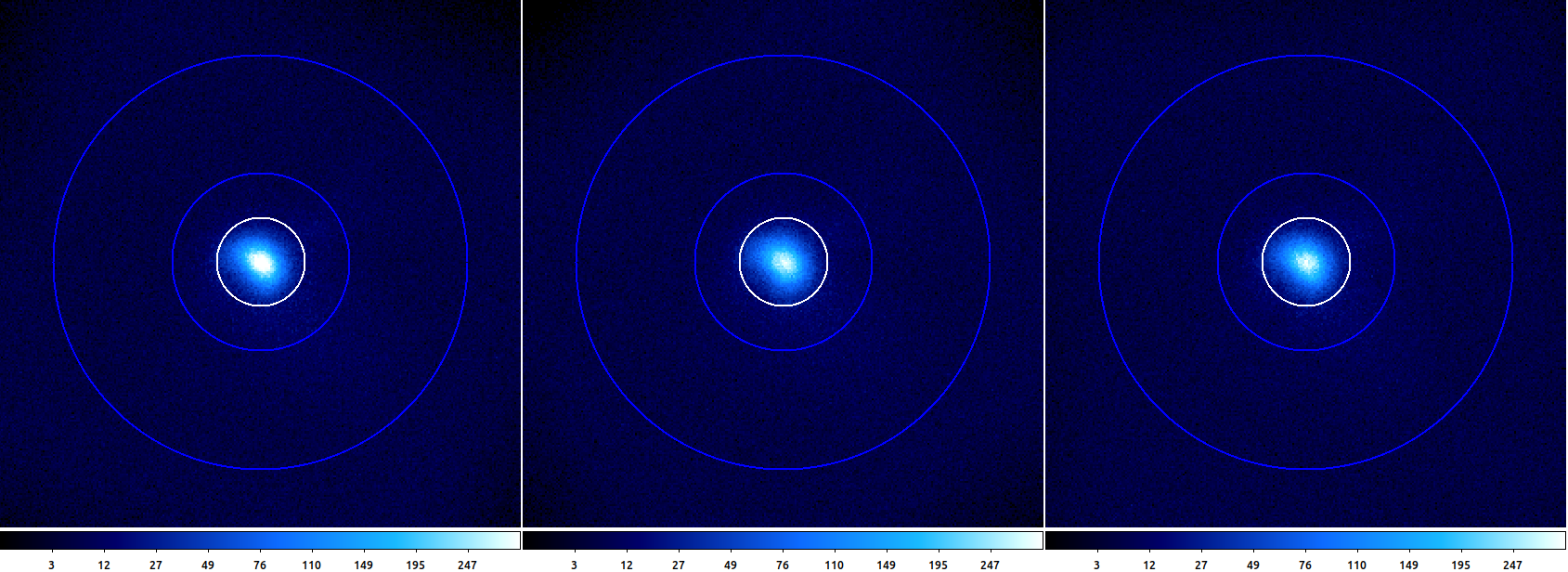}
\caption{Total nebula source (white 1$^{\prime}$ radius circle) and background (blue annulus, inner radius 2$^{\prime}$, outer radius 4.7$^{\prime}$) regions shown on images from DU1 (left) to DU3 (right). Intensity is on a logarithmic scale to bring out the faint background.}
\label{fig:bkg_sel}
\end{figure}

The field of view (FoV) of each detector is 12.9$^{\prime}\times$ 12.9$^{\prime}$ (with the three detectors mounted at angles of 120$^\circ$ to each other) \cite{2022JATIS...8b6002W}\!, while the X-ray emission of Vela PWN covers a much smaller region \cite{2001ApJ...554L.189P}\!. We identified source and background regions using the SAOImageDS9 software \cite{2006ASPC..351..574J}\!. The source region is a circle of radius 1$^{\prime}$ centered on the pulsar. The background region is an annulus with inner radius of 2$^{\prime}$ and outer radius of 4.7$^{\prime}$ (see Extended Data Fig.\,\ref{fig:bkg_sel}).
 
The polarization data consists of Stokes parameters ($I$, $Q$, $U$) for each photon in the event\_l2 files. For details on the Stokes parameters as defined in IXPE data see ref. \cite{Di_Marco_2022}.
Data analysis was performed with different techniques and by different groups within the IXPE collaboration to cross-check the results. We first analyzed the data with the IXPE internal software \textit{ixpeobssim} \cite{2022arXiv220306384B}\!, customized for IXPE data analysis and simulations. \textit{ixpeobssim} includes tools developed to determine the source polarization properties via an analysis independent of the spectral models, following the unbinned method described in ref. \cite{2015APh....68...45K}\!.
A second reduction is performed with \textsc{xspec}, using the version \textsc{heasoft} 6.30.1, which includes models for spectro-polarimetric analysis \cite{2017ApJ...838...72S}\!. 
With the binned spectra of the Stokes parameters, \textsc{xspec} performs standard forward-fitting and derives model-dependent polarization parameters. 
The adopted IXPE response functions are from the HEASARC IXPE CALDB March 14$^{th}$, 2022 release.

\begin{figure}
\renewcommand{\figurename}{Extended Data Fig.} 
\centering
\includegraphics[width=1.0\textwidth]{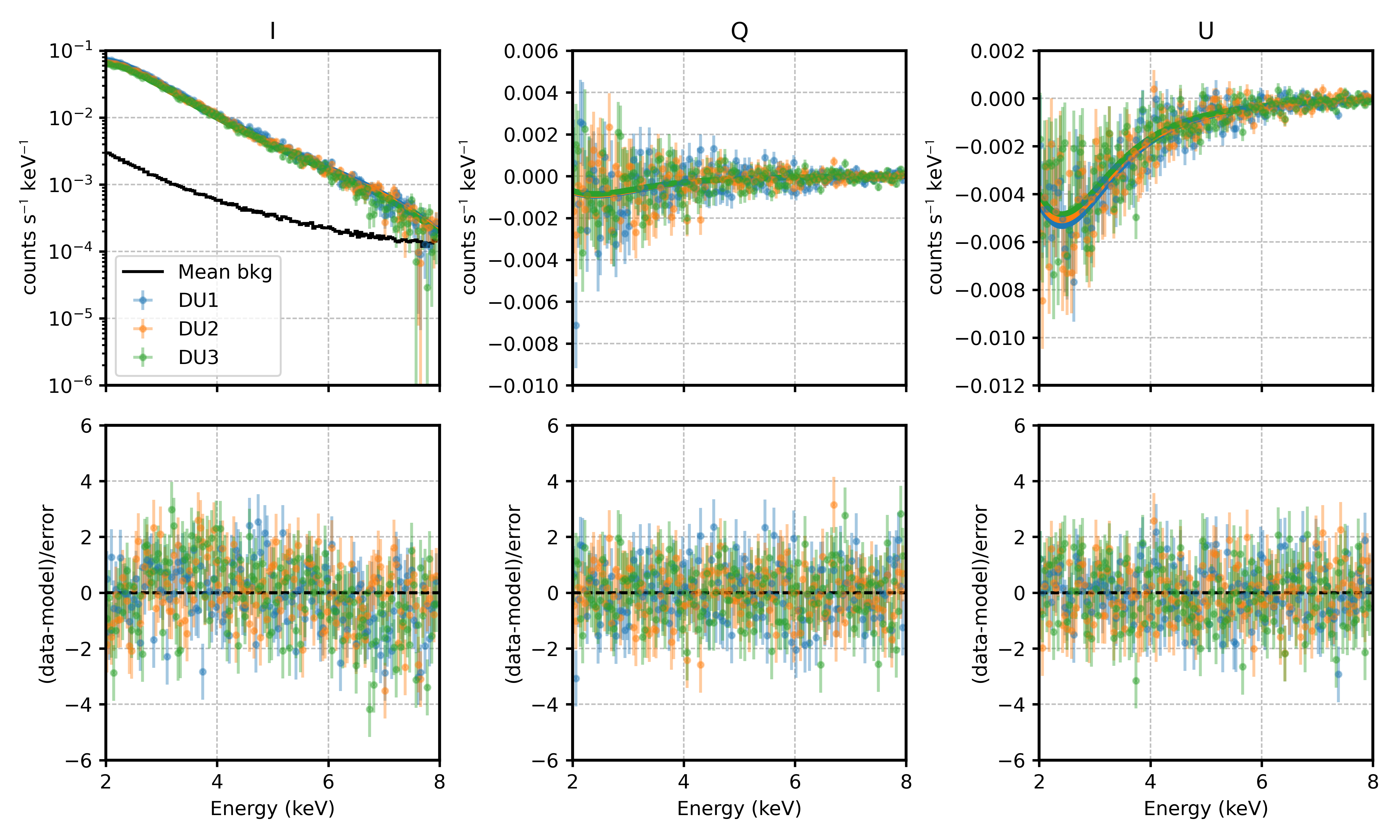}
\caption{Spectral joint fitting for the Stokes parameters in the 2--8\,keV energy band for the three DUs using model \textsc{tbabs*powerlaw*polconst} with previously fit spectral parameters fixed. The average background count $I$ spectrum for the three DUs is shown in black. Fit residuals are shown at the bottom.}
\label{fig:stokes_spectrum}
\end{figure}

The spectral analysis shows that background events are not negligible at high energies, with rates becoming comparable (see Extended Data Fig.\,\ref{fig:stokes_spectrum}). To determine the source polarization while avoiding dilution effects at high energies, the background component was subtracted in all the analyses. The 2--8\,keV band-averaged, aperture-averaged (1$^{\prime}$ radius) polarization degree and angle measurements are summarized in Extended Data Table\,\ref{tab:spec_pol_energy_bin}. The results are consistent between the three DUs and the two different analysis methods.

Spectro-polarimetric analysis is performed using \textsc{xspec} \cite{1996ASPC..101...17A} to jointly fit the three DUs in a two-step procedure. In the first step, the $I$ energy distribution are fitted with a spectral model.  In the second step the spectral model is fixed, while $U$ and $Q$ are fit. This method thus does a forward folding fit to the binned spectra of the Stokes parameters $I$, $Q$, and $U$ jointly after fixing the spectral model. 

The Vela PWN binned $I$ spectra from the three DUs were fitted with the model \textsc{factor*tbabs*powerlaw}, where \textsc{tbabs} takes into account the interstellar absorption -- here, we fixed the column density \cite{2001ApJ...552L.129P} to $N_H=0.03\times10^{22}$ cm$^{-2}$ -- a relative normalization \textsc{factor} for DU2 and DU3 accounts for uncertainties in their absolute effective area. 
The best-fit curves and the spectra of the Vela PWN for the three DUs are shown in Extended Data Fig.\,\ref{fig:stokes_spectrum}. At present, uncertainties in the effective area energy dependence are substantial, due to the incomplete knowledge of telescope alignment and related vignetting. Accordingly we focus on the polarization information. The spectral model parameters do not impact the polarization results, which are affected only by the energy assigned to individual photons. The individual photon energy estimate is tied to frequent in-flight calibration, performed with onboard calibration source during the Earth occultation. As the effective area variation becomes better understood, we expect the spectral model fitting to improve.

With the spectral model, binned $I$, $Q$, and $U$ spectra are fitted to obtain the polarimetric information, using the polarization model \textsc{polconst} of \textsc{xspec}. 
The fitted $I$, $Q$, $U$ spectra in the 2--8\,keV energy band are shown in Extended Data Fig.\,\ref{fig:stokes_spectrum}. The polarization angle and degree obtained are in good agreement with those obtained by the model-independent analysis performed with \textit{ixpeobssim}, as reported in Extended Data Table\,\ref{tab:spec_pol_energy_bin}.

\begin{figure}
\renewcommand{\figurename}{Extended Data Fig.} 
\centering
\includegraphics[width=1.0\textwidth]{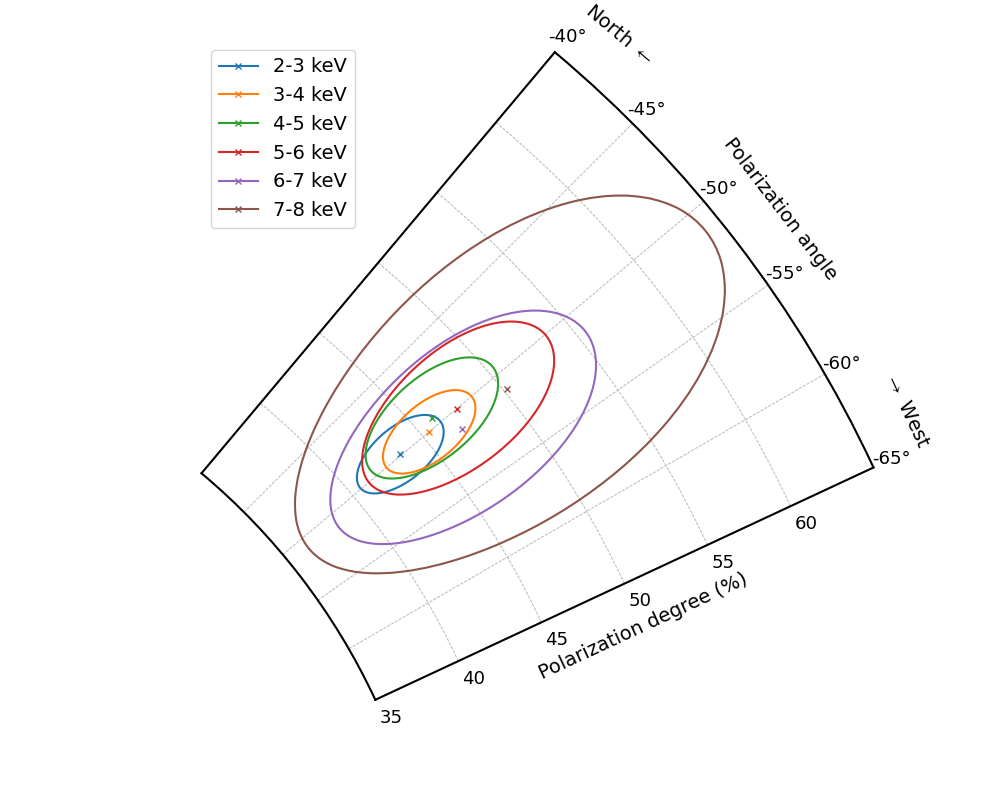}
\caption{Polar plot showing the polarization degree (PD) and polarization angle (PA) fit with data from the three DUs, for different energy bands. Ellipses show the 68.3\% confidence level errors obtained with \textsc{xspec}. 
}
\label{fig:polar_cont_energy}
\end{figure}

\begin{figure}
\renewcommand{\figurename}{Extended Data Fig.} 
\centering
\includegraphics[width=0.8\textwidth]{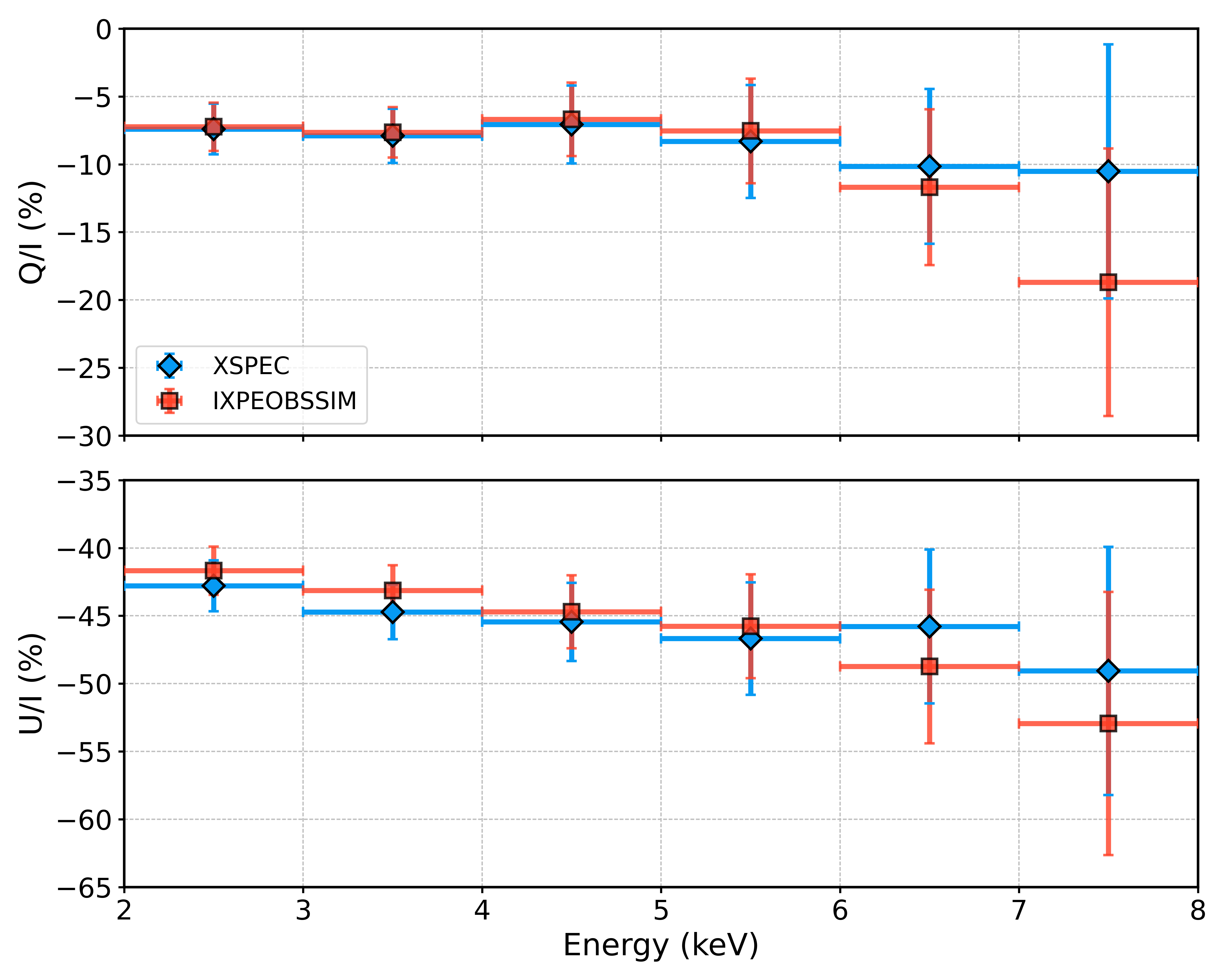}
\caption{$Q/I$ (top) and $U/I$ (bottom) Stokes parameters of the Vela PWN as functions of energy in 2--8\,keV range derived with \textit{ixpeobssim} and \textsc{xspec} independently. The results are from the joint analysis of three DUs.
}
\label{fig:q_uVSenergy}
\end{figure}

The same procedure was used to investigate the polarization in six energy bins fitting with \textsc{xspec} model \textsc{factor*tbabs*powerlaw*polconst} in each energy interval, freezing the $I$ spectral model parameters to the values previously obtained from the 2--8\,keV analysis. Polarization degree and angle in each energy bin are reported in Extended Data Table\,\ref{tab:spec_pol_energy_bin} and Fig.\,\ref{fig:pol_energy}. The \textit{ixpeobssim} analysis is fully in agreement with that of \textsc{xspec}. PD and PA values obtained with \textsc{xspec} are represented in a 2D polar plot -- with contours enclosing the 68.3\% C.L. regions -- in Extended Data Fig.\,\ref{fig:polar_cont_energy}. Similarly, $Q/I$ and $U/I$ Stokes parameters obtained with \textsc{xspec} and \textit{ixpeobssim} are shown in Extended Data Fig.\,\ref{fig:q_uVSenergy}. 

We have attempted to quantify the significance of the PD$(E)$ trend in Fig.\,\ref{fig:pol_energy}. While a $\chi^2$ test finds that a constant PD is acceptable at the 57\% level, a run-test analysis only accepts this hypothesis at 7.7\% and a Kolmogorov-Smirnov test only has a probability of 3.6\%. In contrast a linear PD$(E)$ model gives an acceptable $\chi^2$ probability at the 98\% level, the run-test statistic is allowed at 41\%, and a K-S test accepts the hypothesis at the 98\% level. Thus a linear model with slope (1.8$\pm$0.9)$\times 10^{-2}$ keV$^{-1}$, while not demanded by the data, is preferred over a simple constant PD model at the 2$\sigma$ confidence level.

\subsubsection*{Spatially-resolved Analysis}
Spatially-resolved analysis is performed with \textit{ixpeobssim}. We divided the FoV into 30$^{\prime\prime}$ square bins, guided by the mirror PSF. Polarization analysis was performed for each bin, without background subtraction. Removing the background counts would give a small increase in the PD without affecting the PA.
The statistics in the outer bins are poor, so we present only the results in 5$\times$5 grids centered on the pulsar (Fig.\,\ref{fig:pol_map}). The tabulated results of the normalized $Q$, $U$, and their significance are in Extended Data Table\,\ref{tab:QU_map_3dus}, and the corresponding PD and PA are presented in Extended Data Table\,\ref{tab:pol_map_3dus}.

\begin{figure}
\renewcommand{\figurename}{Extended Data Fig.} 
\centering
\includegraphics[width=0.7\textwidth]{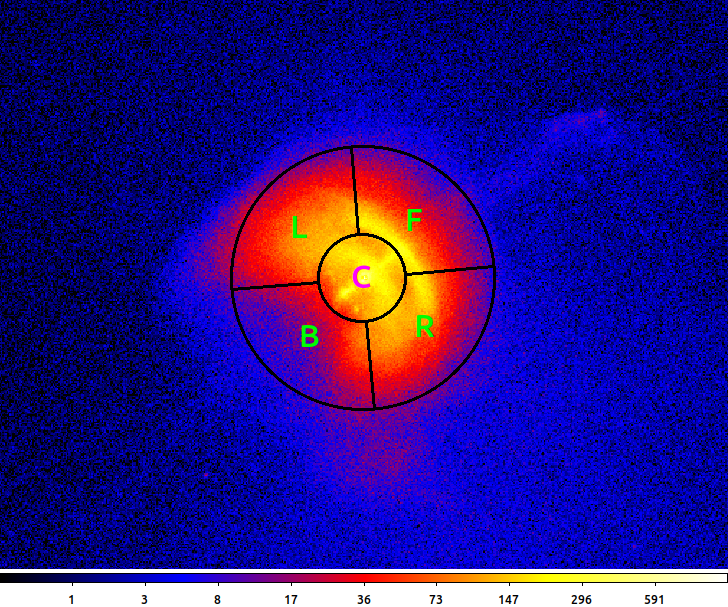}
\caption{An alternative spatial partition of the PWN with regions aligned with the PWN symmetry axis on image observed by Chandra. $L$, $R$, $F$, $B$ label the left, right, front, back regions with respect to the $C$ center region respectively, corresponding to the analysis tabulated in Extended Data Table\,\ref{tab:panda}. The inner circle has a radius of 15$^{\prime\prime}$ and the outer circle radius is 45$^{\prime\prime}$.
}
\label{fig:panda_region}
\end{figure}

We also analyze the polarization parameters in five larger regions, aligned to the PWN symmetry axis, labeled as $L$, $R$, $F$, $B$ and $C$ in Extended Data Fig.\,\ref{fig:panda_region}. Here we use \textit{ixpeobssim} and subtract a scaled off-source background. The results, tabulated in Extended Data Table\,\ref{tab:panda}, show a polarization degree increasing along the symmetry axis from $\sim 30\%$ ($B$ region) up to $\sim 70\%$ ($F$ region).
As with the finer square grid, $L$ and $R$ side regions show PA symmetrically spread about PA for the symmetry axis region $C$.

\subsection*{The effect of the Pulsar on the PWN Polarization}

The 89\,ms Vela pulsar's X-rays are largely thermal and make a negligible contribution to the 2--8\,keV PWN flux as a whole. Thus, they cannot significantly affect our image-averaged PD and PA. However, from high resolution Chandra images, we see that the pulsar contributes $\sim 10\%$ of the 2--8\,keV flux in the central square region of Fig.\,\ref{fig:pol_map} and so it might affect its polarization. To bound such effects, we measured the spectrum of the Vela pulsar using the Chandra observation obsid 218, processed through the standard pipeline with the CIAO 4.13 package and CALDB version 4.9.5. If the pulsar contributes an unpolarized background, the corrected nebula-only PD rises to 55\%. If in contrasts the pulsar emission is 100\% polarized parallel (perpendicular) to the PA measured for the central square region, then the pulsar-subtracted polarization falls (rises) to 39\% (66\%). These values are from DU1 measurements only since with the smallest half-power width the pulsar will be most severe for this DU. In practice, the pulsar polarization is quite low in the optical band (PD=8.1$\pm$0.6\% at PA=(146.3$\pm$0.6)$^\circ$) \cite{2014MNRAS.445..835M}\!, and, by analogy with the Crab, will be even lower in the X-rays, so the true effect should be close to the unpolarized case, boosting the inferred PWN central PD. We apply this correction to the central bin in Fig.\,\ref{fig:pol_map}. Follow-up IXPE studies may more tightly bound the polarization of the pulsed emission. 

\section*{Data Availability}
IXPE data are available through the NASA's HEASARC data archive at \url{https://heasarc.gsfc.nasa.gov}. Other derived data, supporting the findings of this study, are available from the corresponding author upon request.

\section*{Code Availability}
The High Energy Astrophysics Science Archive Research Center (HEASARC) developed the HEASOFT (HEASARC Software). We used the HEASOFT version 6.30.1 package for the spectro-polarimetric IXPE data analysis, available at: \url{https://heasarc.gsfc.nasa.gov/docs/software/heasoft/}. The proper instrument response functions are provided by the IXPE Team as a part of the IXPE calibration database released on 2022 March 14 and available in the HEASARC Calibration Database at: \url{https://heasarc.gsfc.nasa.gov/docs/heasarc/caldb/caldb_supported_missions.html}. 

The software developed by IXPE collaboration is available publicly through the web-page \url{https://ixpeobssim.readthedocs.io/en/latest/?badge=latest.}

Information supporting the findings of this study is available from the corresponding author upon request.

\section*{Acknowledgements} 
The Imaging X ray Polarimetry Explorer (IXPE) is a joint US and Italian mission.  The US contribution is supported by the National Aeronautics and Space Administration (NASA) and led and managed by its Marshall Space Flight Center (MSFC), with industry partner Ball Aerospace (contract NNM15AA18C). The Italian contribution is supported by the Italian Space Agency (Agenzia Spaziale Italiana, ASI) through contract ASI-OHBI-2017-12-I.0, agreements ASI-INAF-2017-12-H0 and ASI-INFN-2017.13-H0, and its Space Science Data Center (SSDC) with agreements ASI-INAF-2022-14-HH.0 and ASI-INFN 2021-43-HH.0, and by the Istituto Nazionale di Astrofisica (INAF) and the Istituto Nazionale di Fisica Nucleare (INFN) in Italy. This research used data products provided by the IXPE Team (MSFC, SSDC, INAF, and INFN) and distributed with additional software tools by the High-Energy Astrophysics Science Archive Research Center (HEASARC), at NASA Goddard Space Flight Center (GSFC).
The research at Guangxi University was supported in part by National Natural Science Foundation of China (Grant No. 12133003).
The research at Boston University was supported in part by National Science Foundation grant AST-2108622.
I.A. acknowledges financial support from the Spanish ``Ministerio de Ciencia e Innovaci\'on'' (MCINN) through the ``Center of Excellence Severo Ochoa'' award for the Instituto de Astrof\'isica de Andaluc\'ia-CSIC (SEV-2017-0709) and through grants AYA2016-80889-P and PID2019-107847RB-C44.

\section*{Author contributions} 
F.X. led the data analysis and the writing of the paper. 
A.D.M. and F.L.M performed the spectro-polarimetric analysis and contributed to the manuscript. F.M. and J.R. contributed to data analysis and energy calibration of the data-set.
K.L. and N.B. contributed to the data analysis and results interpretation. 
R.W.R. and L.L. helped revise the manuscript. 
E.C., P.S., M.B., S.F., R.F., M.P. and A.T. contributed to the interpretation of the results and to text revisions. 
N.D.L., S.G., N.O., M.N. and E.W. performed independent analysis of the data. 
The remaining members of the IXPE collaboration contributed to the design of the mission, to the calibration of the instrument, to definition its science case and to the planning of the observations. 
All authors provided inputs and comments on the manuscript.

\section*{Competing interests} 
The authors declare no conflicts of interest. 

\newpage
\section*{Extended Data}
%
\begin{figure}[h]
\renewcommand{\figurename}{Extended Data Fig.} 
\setcounter{figure}{0}
\centering
\includegraphics[width=1.0\textwidth]{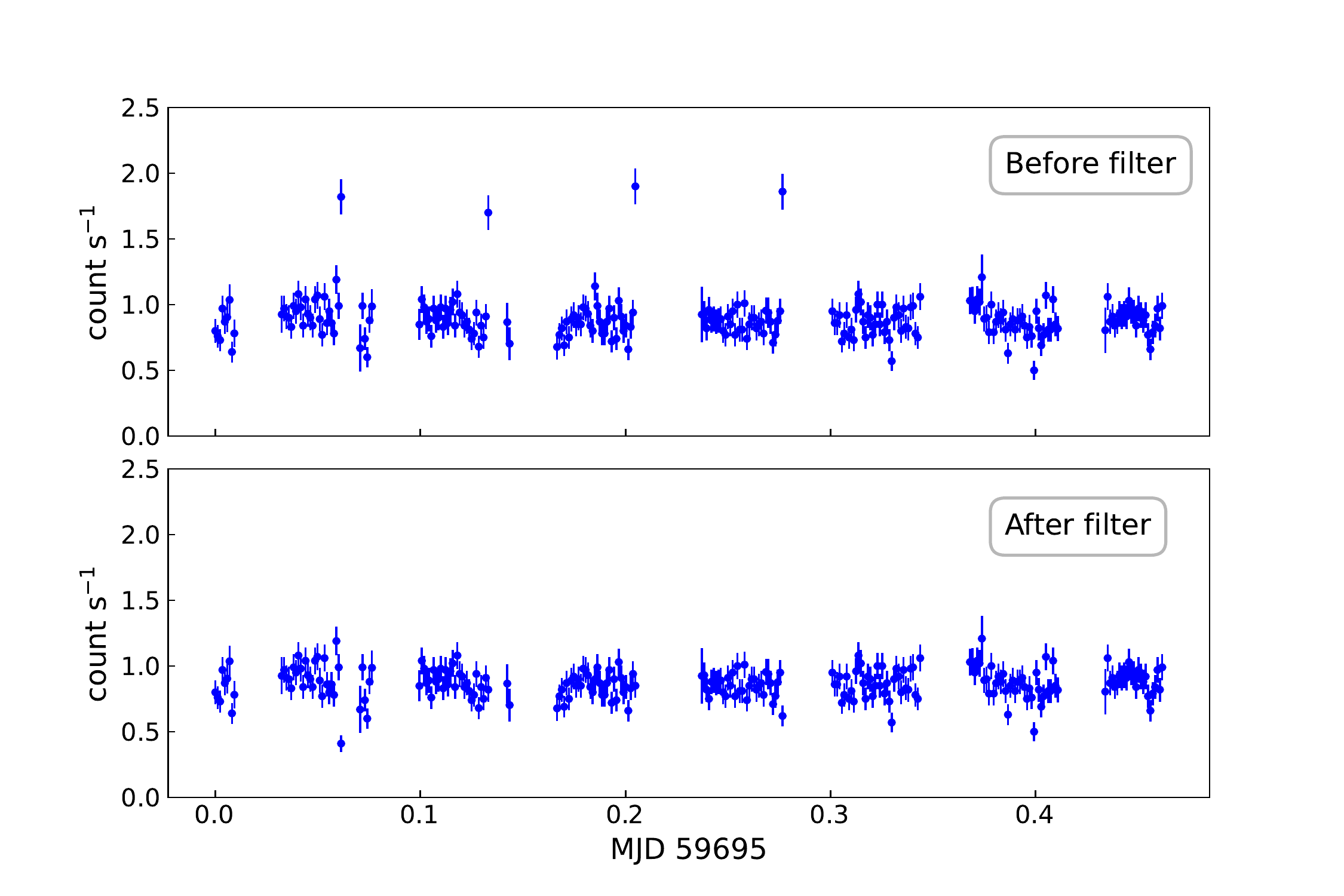}
\caption{Light curve of the Vela PWN observed by IXPE DU1 before and after data filtering (top and bottom panels, respectively). Filtering the high count-rate excursions from the event\_l2 file removes $\sim 1.47$\% of the exposure time.}
\label{fig:solar_remove}
\end{figure}
%

\clearpage
%

\clearpage
%

\clearpage
%

\clearpage
%

\clearpage
%

\clearpage
\begin{table}
\renewcommand{\tablename}{Extended Data Table} 
\centering
\renewcommand\arraystretch{1}{
\begin{threeparttable} 
\caption{\textbf{Polarization degree and angle in six different energy bands and in the full IXPE 2--8\,keV band.}}
\setlength{\tabcolsep}{1mm}{
\begin{tabular}{lcccccc|c}
\toprule
\specialrule{0em}{1pt}{1pt}
{} & \textbf{2--3\,keV} & \textbf{3--4\,keV} & \textbf{4--5\,keV} & \textbf{5--6\,keV} & \textbf{6--7\,keV} & \textbf{7--8\,keV} & \textbf{2--8\,keV}\\
\midrule
PD (\%)$^I$ &   42.3$\pm$1.8 &   43.8$\pm$1.9 &   45.2$\pm$2.7 &    46.4$\pm$3.8 &    50.1$\pm$5.7 &   56.1$\pm$9.7 & 44.6$\pm$1.4 \\
\specialrule{0em}{1pt}{1pt}
PD (\%)$^X$ &   43.4$\pm$1.9 &   45.4$\pm$2.0 &   46.0$\pm$2.9 &    47.4$\pm$4.1 &    46.9$\pm$5.7 &   50.2$\pm$9.2 & 44.9$\pm$1.1 \\
\midrule
PA ($^\circ$)$^I$ & $-$49.9$\pm$1.2 & $-$50.0$\pm$1.2 & $-$49.3$\pm$1.7 & $-$49.7$\pm$2.4 & $-$51.7$\pm$3.3 & $-$54.7$\pm$5.0 & $-$50.0$\pm$0.9 \\
\specialrule{0em}{1pt}{1pt}
PA ($^\circ$)$^X$ & $-$49.9$\pm$1.2 & $-$50.0$\pm$1.3 & $-$49.4$\pm$1.8 & $-$50.0$\pm$2.5 & $-$51.2$\pm$3.5 & $-$51.1$\pm$5.4 & $-$50.0$\pm$0.7 \\
\bottomrule
\label{tab:spec_pol_energy_bin}
\end{tabular}
}
\begin{tablenotes}
\item[]Polarization degree and polarization angle measured using \textit{ixpeobssim} and \textsc{xspec} in six energy bands and 2--8\,keV using three DUs. Uncertainties on the polarization degree and angle are calculated for a 68.3\% confidence level, assuming that they are independent. The values from the two different methods are compatible within the uncertainties.
\item[\textit{I}] Values are obtained with \textit{ixpeobssim}.
\item[\textit{X}] Values are obtained with \textsc{xspec}.
\end{tablenotes}
\end{threeparttable}}
\end{table}

\clearpage
\begin{table}
\renewcommand{\tablename}{Extended Data Table} 
\centering
\renewcommand\arraystretch{1.5} 
\begin{threeparttable} 
\caption{\textbf{The normalized Stokes parameters and the detection significance within the bins of Fig.\,\ref{fig:pol_map}.}}
\begin{tabular}{c|ccccc|c} 
\toprule 
 & \textbf{-2}$^b$ & \textbf{-1}$^b$ & \textbf{0}$^b$ & \textbf{1}$^b$ & \textbf{2}$^b$ & \\
\midrule 
&$0.33\pm$0.17 &$0.20\pm$0.13 &$0.07\pm$0.11 &$-0.09\pm$0.13 &$-0.14\pm$0.15 & Q/I$^c$\\
\textbf{2}$^a$ 
&$-0.18\pm$0.18 &$-0.18\pm$0.13 &$-0.61\pm$0.11 &$-0.36\pm$0.13 &$-0.45\pm$0.14 & U/I$^d$\\
&$1.3$ &$1.2$ &$5.0$ &$2.0$ &$2.6$ & Sig$^e$\\
\midrule 
&$0.32\pm$0.10 &$0.34\pm$0.05 &$0.05\pm$0.04 &$-0.10\pm$0.07 &$-0.21\pm$0.13 & Q/I$^c$\\
\textbf{1}$^a$ 
&$0.07\pm$0.10 &$-0.34\pm$0.05 &$-0.53\pm$0.04 &$-0.56\pm$0.07 &$-0.42\pm$0.13 & U/I$^d$\\
&$2.5$ &$9.7$ &$12.9$ &$7.6$ &$3.1$ & Sig$^e$\\
\midrule 
&$0.10\pm$0.09 &$0.13\pm$0.04 &$-0.09\pm$0.02 &$-0.19\pm$0.04 &$-0.08\pm$0.11 & Q/I$^c$\\
\textbf{0}$^a$ 
&$-0.03\pm$0.09 &$-0.32\pm$0.04 &$-0.48\pm$0.02 &$-0.60\pm$0.04 &$-0.43\pm$0.11 & U/I$^d$\\
&$-$ &$8.8$ &$19.7$ &$15.6$ &$3.3$ & Sig$^e$\\
\midrule 
&$-0.01\pm$0.12 &$-0.20\pm$0.07 &$-0.30\pm$0.04 &$-0.44\pm$0.05 &$-0.18\pm$0.12 & Q/I$^c$\\
\textbf{-1}$^a$ 
&$-0.20\pm$0.12 &$-0.19\pm$0.07 &$-0.53\pm$0.04 &$-0.37\pm$0.05 &$-0.40\pm$0.12 & U/I$^d$\\
&$0.8$ &$3.2$ &$9.6$ &$10.5$ &$3.0$ & Sig$^e$\\
\midrule 
&$-0.07\pm$0.15 &$0.04\pm$0.13 &$-0.35\pm$0.09 &$-0.42\pm$0.11 &$-0.13\pm$0.13 & Q/I$^c$\\
\textbf{-2}$^a$ 
&$-0.33\pm$0.15 &$-0.01\pm$0.13 &$0.05\pm$0.09 &$-0.09\pm$0.12 &$-0.11\pm$0.13 & U/I$^d$\\
&$1.4$ &$-$ &$3.1$ &$3.2$ &$0.1$ & Sig$^e$\\
\bottomrule
\end{tabular}
\label{tab:QU_map_3dus}
\begin{tablenotes}
\item[\textit{a}] The row number where 0 presents the center bin containing the pulsar.
\item[\textit{b}] The column number where 0 presents the center bin containing the pulsar.
\item[\textit{c}] The $Q$/$I$ Stokes parameters with 68.3\% confidence level error.
\item[\textit{d}] The $U$/$I$ Stokes parameters with 68.3\% confidence level error.
\item[\textit{e}] The significance ($\sigma$) for the non-polarized hypothesis test, based on the standard normal distribution. 
The corresponding P value means the probability that a polarized signal is generated by a non-polarized source. Specifically, a significance of 2$\sigma$ means that the probability that the emission is polarized reaches 95\%. Conventionally, the minimum detectable polarization at 99\% confidence level (MDP$_{99}$) corresponds to a significance of 2.33$\sigma$. 
\item[$-$] No significant polarization detection. 
\end{tablenotes}
\end{threeparttable}
\end{table}

\clearpage
\begin{table}
\renewcommand{\tablename}{Extended Data Table} 
\centering
\renewcommand\arraystretch{1.5} 
\begin{threeparttable} 
\caption{\textbf{The Polarization degree and angle within the bins of Fig\,\ref{fig:pol_map}.}}
\begin{tabular}{c|ccccc|c} 
\toprule 
 & \textbf{-2}$^b$ & \textbf{-1}$^b$ & \textbf{0}$^b$ & \textbf{1}$^b$ & \textbf{2}$^b$ & \\
\midrule
\multirow{2}{*}{\textbf{2}$^a$} & 37$\pm$18 & 27$\pm$13 &61$\pm$12 &37$\pm$13 &47$\pm$15  & PD$^c$\\
&$-14\pm$14 &$-21\pm$14 &$-41.7\pm$5.3 &$-52\pm$10 &$-53.8\pm$8.9       & PA$^d$\\
\midrule
\multirow{2}{*}{\textbf{1}$^a$} &33$\pm$10 &48.5$\pm$5.0 &53.5$\pm$4.1 &56.8$\pm$7.1 &47$\pm$13   & PD$^c$\\ 
&6.3$\pm$9.0 &$-22.4\pm$3.0 &$-42.2\pm$2.2 &$-50.2\pm$3.6 &$-58.2\pm$7.7    & PA$^d$\\
\midrule
\multirow{2}{*}{\textbf{0}$^a$} &$10.3\pm$8.8 &34.4$\pm$3.9 &49.0$\pm$2.5 &62.8$\pm$4.0 &44$\pm$11   & PD$^c$\\
& $-$7.4$\pm$24  &$-34.3\pm$3.3 &$-50.3\pm$1.5 &$-53.9\pm$1.9 &$-50.5\pm$7.4      & PA$^d$\\
\midrule
\multirow{2}{*}{\textbf{-1}$^a$} &21$\pm$12 &27.5$\pm$7.2 &38.5$\pm$4.0 &57.1$\pm$5.4 &44$\pm$12   & PD$^c$\\
&$-47\pm$17 &$-68.3\pm$7.5 &$-70.0\pm$3.0 &$-69.8\pm$2.7 &$-57.3\pm$7.9     & PA$^d$\\
\midrule
\multirow{2}{*}{\textbf{-2}$^a$} &34$\pm$15 &$4.5^{+13}_{-4.5}$ &34.9$\pm$9.5 &43$\pm$12 &17$\pm$14   & PD$^c$\\
& $-51\pm$13 & $-6.0\pm$85  & $86.1\pm$7.8 & $-84.2\pm$7.6 &$-70\pm$23    & PA$^d$\\
\bottomrule
\end{tabular}
\label{tab:pol_map_3dus}
\begin{tablenotes}
\item[]Polarization degree and polarization angle measured using \textit{ixpeobssim}. Uncertainties on the polarization degree and angle are calculated for a 68.3\% confidence level, assuming that they are independent.
\item[\textit{a}] The row number where 0 presents the center bin containing the pulsar.
\item[\textit{b}] The column number where 0 presents the center bin containing the pulsar.
\item[\textit{c}] The polarization degree (\%) with 68.3\% confidence level error.
\item[\textit{d}] The polarization angel ($^{\circ}$) with 68.3\% confidence level error.

\end{tablenotes}
\end{threeparttable}
\end{table}

\clearpage
\begin{table}
\renewcommand{\tablename}{Extended Data Table} 
\centering
\renewcommand\arraystretch{1.5} 
\begin{threeparttable} 
\caption{\textbf{The polarization degree and angle in the five regions shown in Extended Data Fig.\,\ref{fig:panda_region}.}}
\begin{tabular}{lccccc}
\toprule
{} & \textbf{Center} & \textbf{Front} & \textbf{Right}  & \textbf{Left}  & \textbf{Back} \\
{} & \textbf{[ C ]} & \textbf{[ F ]} & \textbf{[ R ]}  & \textbf{[ L ]}  & \textbf{[ B ]} \\
\midrule
PD (\%)$^S$ &   49.6$\pm$2.5 &   70.0$\pm$3.6 &   56.0$\pm$3.1 &    42.0$\pm$3.0 &    33.3$\pm$3.6\\
\specialrule{0em}{1pt}{1pt}
PD (\%)$^{S+B}$ &   48.8$\pm$2.5 &   65.4$\pm$3.6 &   53.1$\pm$3.1 &    39.9$\pm$3.0 &    31.1$\pm$3.7\\
\midrule
PA ($^\circ$)$^S$ &  $-$50.2$\pm$1.5 &  $-$48.8$\pm$1.5 &  $-$64.9$\pm$1.6 &   $-$30.1$\pm$2.1 &   $-$59.4$\pm$3.1\\
\specialrule{0em}{1pt}{1pt}
PA ($^\circ$)$^{S+B}$ &  $-$50.3$\pm$1.5 &  $-$48.8$\pm$1.6 &  $-$64.9$\pm$1.7 &   $-$30.2$\pm$2.2 &   $-$59.4$\pm$3.4\\
\bottomrule
\end{tabular}
\label{tab:panda}
\begin{tablenotes}
\item[]Polarization degree and polarization angle in C, F, R, L, and B regions measured using \textit{ixpeobssim} with and without background subtraction. Background regions are the same as shown in Extended Data Fig.\,\ref{fig:bkg_sel}. Uncertainties on the polarization degree and angle are calculated for a 68.3\% confidence level, assuming that they are independent.
\item[$^S$] Values are obtained with background subtraction.
\item[$^{S+B}$] Values are obtained without background subtraction.
\end{tablenotes}
\end{threeparttable}
\end{table}
\clearpage


\end{document}